\documentstyle[multicol,psfig,aps,prl]{revtex}
\draft

\newcommand{\ba}{\begin{eqnarray}}                                             
\newcommand{\ea}{\end{eqnarray}}     

\newcommand{\be}{\begin{equation}}                                             
\newcommand{\ee}{\end{equation}}     
\begin{document}

\title{Ultrafast Spin Dynamics in Nickel}
\author{W. H\"ubner$^{1, 2}$ and G. P. Zhang$^2$}
\address{$^1$ Institut de Physique et Chimie des Materiaux de Strasbourg,
Unite Mixte 380046 CNRS-ULP-EHICS,\\
23, rue du Loess, 67037 Strasbourg Cedex, France\\ 
$^2$Max-Planck-Institut f\"ur Mikrostrukturphysik, Weinberg 2,
D-06120 Halle, Germany$^*$}

\date{\today}
\maketitle

\begin{abstract}
The spin dynamics in Ni
is studied 
by an exact diagonalization method on the ultrafast time scale. It
is shown that the femtosecond relaxation of the magneto-optical response results
from 
exchange interaction and spin-orbit coupling. Each of the two mechanisms affects
 the relaxation
process differently. We find that the intrinsic spin dynamics occurs during
about 10 fs while extrinsic effects such as laser-pulse duration and spectral
width can slow down the observed dynamics considerably. 
Thus, our theory indicates that there is still room to
accelerate the spin dynamics in experiments. 
\end{abstract}
\begin{multicols}{2}

The potential  application of ferromagnetic
  materials on ultrafast time scales is 
attractive for information  storage.  Both experimentally and theoretically the
ultrashort time  behavior
of spin dynamics in transition metals is a new and challenging area.
Vaterlaus {\it et al.}~\cite{vbm} were the first to study the spin dynamics in 
ferromagnetic Gd. Employing spin- and time- resolved photo-emission with 60 ps
  probe pulses they found a spin-lattice relaxation (SLR) of
  100$\pm$80 ps.
Using femtosecond optical and magneto-optical pump-probe techniques,  Beaurepaire {\it et
al}\cite{bmdb} 
have
studied the  relaxation  processes of electrons and spins in  ferromagnetic Ni.  They
reported 
that the  magnetization  of a 22 nm thick film drops rapidly during
  the
 first  picosecond and reaches  its minimum after 2 ps.
  Recently,
  by time-resolved second harmonic generation (SHG),
Hohlfeld 
{\it et al.}~\cite{hmkb} found  that even when  electrons  and  lattice  have
  not  reached a common  thermal
equilibrium, the classical $M(T)$ curve can be reproduced for delay times longer than
the  electron  thermalization  time  of  about  $280$ fs.  On the other hand,
 the  transient  magnetization  reaches  its  minimum  $\approx
50$ fs before
electron  thermalization.  Both groups
  used  polycrystalline Ni but different pulse durations: 60 fs~\cite{bmdb} 
vs 150 fs~\cite{hmkb}. Recently even faster spin decays have been observed~\cite{ae}.

At present, not even the mechanism 
  for this ultrafast spin relaxation is known. 
Moreover, it is of great importance to know
whether 
these results already reflect the intrinsic spin relaxation time scale or not.
Theoretically,  even the {\it static} ferromagnetism in transition metals has been a challenging
topic as the electron  correlation is very strong in these systems~\cite{f}.  
The theoretical  treatment of the spin {\it dynamics} is 
limited. 
On the longer time scales, SLR been studied
previously~\cite{hb}, and the theory yielded a relaxation time of 48 ps for Gd,
  in good agreement with the above mentioned experiment~\cite{vbm}.
On this time scale, the main contribution results from anisotropic
  phonon-magnon interaction.
To our knowledge, so far no theoretical
study has been performed about  the spin
dynamics of transition metals  on the {\it femtosecond} time scale, which is apparently
needed.

For the theoretical description of ultrafast nonequilibrium charge and spin
dynamics, one can either rely on the Baym-Kadanoff-Keldysh Green's function
approach~\cite{snl} or employing an exact diagonalization framework. In this Letter, we
prefer the latter method, which does not involve perturbation theory. Thus it
is more suitable to optical excitations far from equilibrium, especially in
the presence of strong electron correlations.
We start from a  spin-independent  single-electron Ni bandstructure (monolayer to mimic a thin film geometry)  and  use an intra-atomic  electron-electron  interaction.  As
shown 
previously~\cite{hf}, to ensure that the interaction  possesses the correct
atomic symmetry, 
it is  necessary  to go  beyond  the  simplest  and most  commonly 
 used form~\cite{os} 
by 
including general  contributions from four different indices.  We also take into
account  spin-orbit coupling (SOC) as an important effect~\cite{bdd}.  
Then the total Hamiltonian reads
\ba
H&=&\sum_{i,j,k,l,\sigma,\sigma',\sigma'',\sigma'''}U_{i\sigma,j\sigma',l\sigma''',k\sigma''}
c_{i\sigma}^{\dag}{c}_{j\sigma'}^{\dag}c_{k\sigma''}{c}_{l\sigma'''}\nonumber\\
&&+\sum_{\nu,\sigma,K}{\cal E}_{\nu}(K)n_{\nu\sigma}(K)+H_{SO} 
\ea
\noindent where $U_{i\sigma,j\sigma',l\sigma''',k\sigma''}$ is the electron
interaction, which can be described in full generality by the three
parameters Coulomb repulsion $U$, exchange interaction $J$, and the exchange
anisotropy $\Delta J$, for the details see
Ref. \cite{hf}. $c_{i\sigma}^{\dag}$ (${c}_{i\sigma}$) are the usual creation
(annihilation) operators in the orbital $i$ with spin $\sigma$
($\sigma=\uparrow,\downarrow$). ${\cal E}_{\nu}(K)$ is the single-particle 
energy spectrum for band $\nu$. $n_{\nu\sigma}(K)$ is the particle number
operator in $K$-space. 
$H_{SO}$ is the
spin orbit coupling~\cite{tbf}. Such kind of Hamiltonian is general
enough to address  the spin dynamics on the ultrafast time scale as it
contains the  necessary  ingredients, such as the electronic Coulomb interaction, exchange
interaction, and the nature of the bands.  However it is not possible to solve it without
further approximation.
Fortunately, in the magneto-optical process, vertical (momentum conserving) 
transitions dominate the
whole spectra.  The momentum  crossing terms in the electron interaction
are  less important, especially on the ultrafast time scale.  For simplicity, 
we ignore those
terms, which are off-diagonal in $K$, 
but keep all the off-diagonal terms of the orbitals in  position space.  The single-particle band structure and spin-orbit  coupling parts are treated  exactly.
Under the above approximation, we are able to solve the Hamiltonian by the
exact diagonalization scheme.
 
The parameters affecting femtosecond spin dynamics fall in two classes:
intrinsic (material specific) and extrinsic (experiment specific). Intrinsic
parameters are: (i)  
Coulomb interaction $U$, (ii) exchange interaction $J$, (iii) exchange 
anisotropy $\Delta J$,  (iv) SOC $\lambda$, and (v) band structure ${\cal
E}(K)$. Extrinsic parameters 
include: (vi) the
photon frequencies for the pump and probe pulses, (vii) different optical
techniques such as pump-probe spectroscopy of reflectivity  and
magneto-optics, SHG, or two-photon photoemission (TPPE), (viii) flux of the pulse,
(ix) laser spectral width, and (x) optical pulse duration. 
For a given sample, 
one can vary these external parameters to actively tune the spin dynamics
rather than to only passively observe it.
In this Letter we focus on the effects of (ii),  (iv), (v), and (ix).

Experimentally when the system is pumped,  the  initial 
 distribution  of states is formed.  We
populate the states according to a Gaussian  distribution,
which mimics the real experimental pulse.  The center of the
 populated states is 
around 2 eV above the ground  state.  The initial  state will  evolve with
 time 
 according to Schr\"odinger's equation. We track the relaxation at 2 eV.

For the charge and spin dynamics,  the response  functions are different.
The diagonal element $|\chi_{zz}^{(1)}|$ of the optical susceptibility
 mainly  reflects  the  contribution  from the charge
dynamics  while  $|\chi_{xy}^{(1)}|$  mostly  
reflects the contribution from the spin  dynamics.  With the help of those 
two
functions, we are able 
to address the different characters of the charge and spin dynamics
separately.

First we switch off spin-orbit coupling. Then $\rm \{S^2, S_z \}$ are good
quantum numbers. Before we go further, it is worth checking whether our
Hamiltonian reasonably describes transition metals such as nickel. Firstly, 
the
band structure is correctly reproduced. Secondly the atomic symmetry is well
preserved, yielding the correct degeneracies.  Thirdly, with 
nonzero Coulomb interaction $U$ and exchange interaction $J$, 
the ground state is  ferromagnetic~\cite{add}, which is  consistent with the
ferromagnetism of the Ni thin film.
However, for $U=J=0$, the  ground  state is a singlet.  
It is interesting to note that the  ferromagnetism  exclusively  results  from the
Coulomb and exchange interactions.  This is a nontrivial result.

In the following we monitor both charge and spin dynamics on the fs
time scale and investigate the influence of different intrinsic and extrinsic
parameters. We start with the generic set of parameters, which is $U_0=12$
eV, $J_0=0.99$ eV, and  $(\Delta J)_0=0.12$ eV as given by
the spectroscopic data for Ni. The band structure is parameterized in the usual
tight-binding form~\cite{wh}. The Gaussian width is taken as broad as 20 eV in 
order to reveal the intrinsic charge and spin responses. In
Figs. 1 (b) and (d),
$|\chi_{xy}^{(1)}(\omega,t)|$ and $|\chi_{zz}^{(1)}(\omega,t)|$, as measured
by typical pump-probe experiments, are shown, which
represent the spin and charge dynamics, respectively. 
$\omega=2$ eV hereafter.
This basic result indicates already a much faster charge and spin dynamics 
than seen in
all existing experiments so far. Consequently, there is room to accelerate
both charge and in particular spin dynamics in experiments.   The second
important result is that the spin dynamics lags behind the charge dynamics by
1 fs, which is an appreciable effect on a time scale of 3 fs. This result is very important for possible applications in magnetic 
storage technology, as it guarantees a separate non-equilibrium  spin
memory time.
 We note in passing that, at no stage of our
calculation,
we had to invoke the notion of either electron or spin
temperature. Particularly the concept of spin temperature is questionable not
only due to the heavy non-equilibrium, but also due to the absence of any
well-defined quasi-particle statistics for the spins.

In order to pinpoint the origin of the spin dynamics, we first vary the exchange
interaction while 
the Coulomb  interaction  $U$ is
fixed at $12$ eV.  
For  reduced $J=J_0/10$
(Figs. 1 (a) and (c)),  one can see a more clearly different behavior
 between  spin  and charge
dynamics.
For $|\chi_{xy}^{(1)}(\omega,t)|$, the main peak is much 
broader than for $|\chi_{zz}^{(1)}(\omega,t)|$.
With  increasing  $J$, the spin  dynamics  begins earlier,  
but still lags behind the
charge  dynamics while the latter is virtually unaffected by the variation of
$J$ (see Figs. 1 (c) and (d)).
  $|\chi_{xy}^{(1)}(\omega,t)|$ always reaches its maximum after
$|\chi_{zz}^{(1)}(\omega,t)|$. 
 An onset of this effect has already been  found in
\cite{hf}. 
The exchange
interaction does  not only affect 
the position of the maximum, but also its subsequent decay:
with the decrease of $J$ from $J_0$ to $J_0/10$
the relaxation  time for spin dynamics increases from 2.2 to 3.4 fs.
(see
Figs. 1 (a) and (b)).

Our  calculations show
that 
the relaxation time can be changed by tuning the exchange strength.  Physically
ferromagnetism mainly results from the exchange interaction, but it has been 
 unknown 
how the exchange affects the spin dynamics on the ultrafast time scale. 
Here we clearly see that it accelerates the relaxation:
since in the ferromagnetic
system the energy scales roughly as $J$, the relaxation time scales as
$1/J$. 
Without SOC, the total spin is a good quantum  number, yet the
spin dynamics exclusively results from the loss of the 
quantum coherence due to the
dephasing of the initial state. This occurs on  different time
scales for charge and spin dynamics. Consequently the spin dynamics is 
delayed as compared to the charge dynamics
due to the exchange coupling $J$.

When  the 
spin-orbit  coupling $\lambda$  is turned on to its generic value
$\lambda_0=0.07$ eV, the relaxation time of 
spin  dynamics is determined by both $\lambda$ and $J$. 
To see the effect of SOC on the relaxation process more clearly,  we set $J=\Delta J=0$ eV and 
choose $\lambda=$ 
 0.07, 1.0  eV. Fig. 2 shows that the
relaxation time becomes shorter if $\lambda$ is larger  while the main peak of
the spectrum  becomes narrower. 
Thus for noble metals, such as gold, or rare earths, where SOC
 is much larger than in Ni,
optical alignment could favorably make use of this
enhanced SOC and generate an ultrafast spin dynamics in TPPE  
in this way even from nonmagnetic metals~\cite{zblp}.

Next we study how bandstructure changes spin and charge dynamics to
demonstrate its material sensitivity. 
 We change the bandstructure  multiplying
all the hopping integrals by a factor of 0.1. 
A smaller hopping integral corresponds to a more atom-like material. 
Here $\{A_0\}$ stand for  the
original hopping integrals for Ni~\cite{wh}. 
Figs. 3 (a)
and  (b) show  the spin and charge dynamics, respectively.
Comparing Figs. 1(b) and (d) with the solid curves in Figs. 3 (a) and  (b),
one may note that upon decreasing the hopping integral from $A_0$ to
$A_0/10$, the recurrent features in both $|\chi_{xy}^{(1)}(\omega,t)|$ and
$|\chi_{zz}^{(1)}(\omega,t)|$ are more obvious and 
the relaxation time for the spin dynamics
increases up to
more than 20 fs for $A_0/10$ (note the different abscissa scales). 
Thus a small hopping integral as appearing in  
nanostructured thin films,
islands, clusters, or some
impurities  in the material, slows down the spin dynamics. 
Besides, the 
reduction of the pulse width from 20 eV to 0.2 eV
further
prolongs the relaxation time to 100 fs (long dashed curves in Figs. 3(a) and 3(b)) , which is close to the experimentally observed relaxation
time. So the laser width (spectral and temporal) has a very important impact on the
relaxation time of spin dynamics, which deserves a detailed study.

In order to further investigate this effect as an example for the variation of
the extrinsic parameters, 
we choose two different laser spectral widths, namely $W=$ 20 eV (full curves in
Figs. 4(a) and (b)) and  0.2 eV (long dashed curves). 
The other
parameters are chosen as the generic values of Ni, namely, 
 $J=J_0$, $\Delta J=(\Delta J)_0$, $A=A_0$, $U=U_0$, and $\lambda=\lambda_0$. With
the increase of the width, the relaxation time is prolonged greatly. From
Fig. 4(a), one may notice that for
$W=20$ eV, 
the decay of the spin dynamics is around 3.2 fs; for $W=0.2$ eV, it 
prolongs to  14 fs. 
The pulse-width
dependent relaxation is also obvious for the charge dynamics (see
Fig. 4(b)).  For $W=20$ eV, it decays around 2 fs; 
for $W=0.2$ eV, it lasts up to  13 fs. For  real
applications, the persistence of the slower decay of the 
spin dynamics is important as it sets the
magnetic memory time. 
Thus one can change extrinsic parameters to influence
the spin dynamics even if one does not change material parameters.

In conclusion, starting from a relativistic many-body Hamiltonian, we
  studied the
spin dynamics on the femtosecond scale.
For the intrinsic parameters, 
it is found that the increase of each of
$\lambda$ and $J$ decreases the relaxation time, but the
individual dependence on each of them is different. This ultrafast dynamics
 results
from both the exchange interaction and SOC and does not involve the lattice
\cite{16}.  This is very different from
SLR in Ref. \cite{vbm}. The SLR time in Ni is about 304
ps as calculated from a formalism similar to that applied to Gd
before\cite{hb}, which can be compared with the experimental value of 400 ps 
in Ni~\cite{sbe}. 
 From our 
calculation
it is suggested that the high-speed limit of  spin dynamics is about 
tens of
femtoseconds, which is not yet exhausted by experiments. Thus, in total one
  has to distinguish four different relaxation processes: (a) electronic 
equilibration (1 fs, due to
  electron-electron  interaction); (b) electron-spin relaxation (a
  few fs due to exchange interaction or SOC); (c) electron-lattice
  thermalization ($\approx $ 1 ps, due to electron-phonon coupling); 
(d) SLR ( $\approx$  100 ps due to
  SOC plus anisotropic crystal-field fluctuations).  
The hopping integral also has a very
important effect on spin dynamics. A small hopping integral slows down the
dynamics. This means that e.g. oxides~\cite{kbm}, exhibiting both dispersive bands and
non-dispersive gap states, might be an ideal playground to tune the dynamical
time scale at will, in particular, employing SHG or TPPE. 
For the extrinsic parameters, such as the laser pulse width, 
a small spectral width favors a slow
decay of the spin dynamics. This is important for applications and further
experimental studies.

One of us (W. H.) gratefully acknowledges the hospitality of the IPCMS of
Strasbourg, where a major part of this work has been performed and
stimulating discussions with E. Beaurepaire, J. -Y. Bigot, D. C. Langreth,
J. -C. Merle, and P. Nordlander. This work has been supported by IPCMS
Strasbourg and the TMR
on NOMOKE (ERB-FMRX-CT 96-0015). 

\noindent $^*$ Mailing address.


\psfig{figure=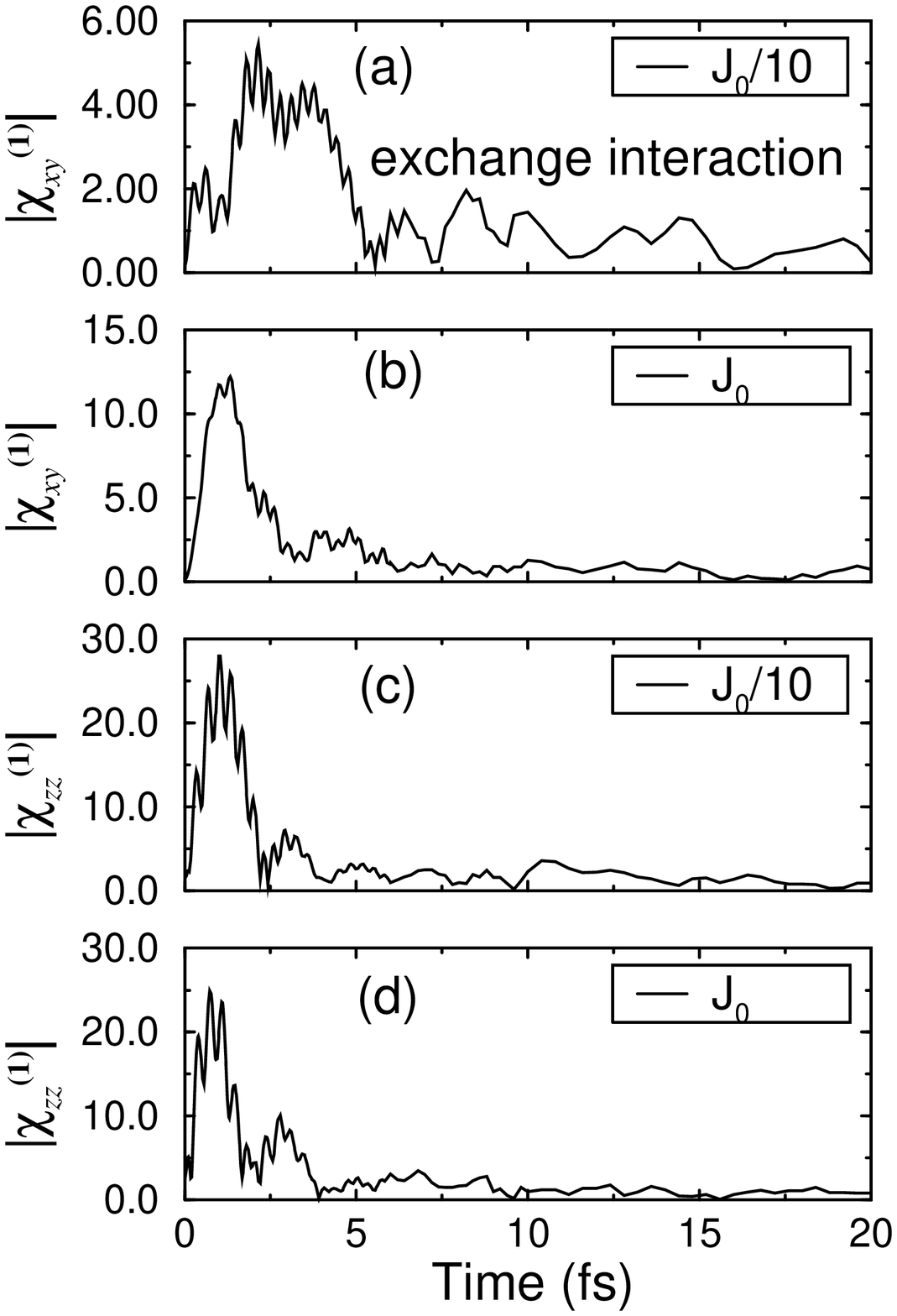,width=8.5cm,angle=0}
\vspace{1.4cm}

{FIG. 1. 
Effect of exchange interaction $J$ ($J=J_0/10$ and $J_0$) 
on spin ((a), (b)) and charge dynamics ((c), (d)). Exchange interaction
dominates the spin decay.}

\psfig{figure=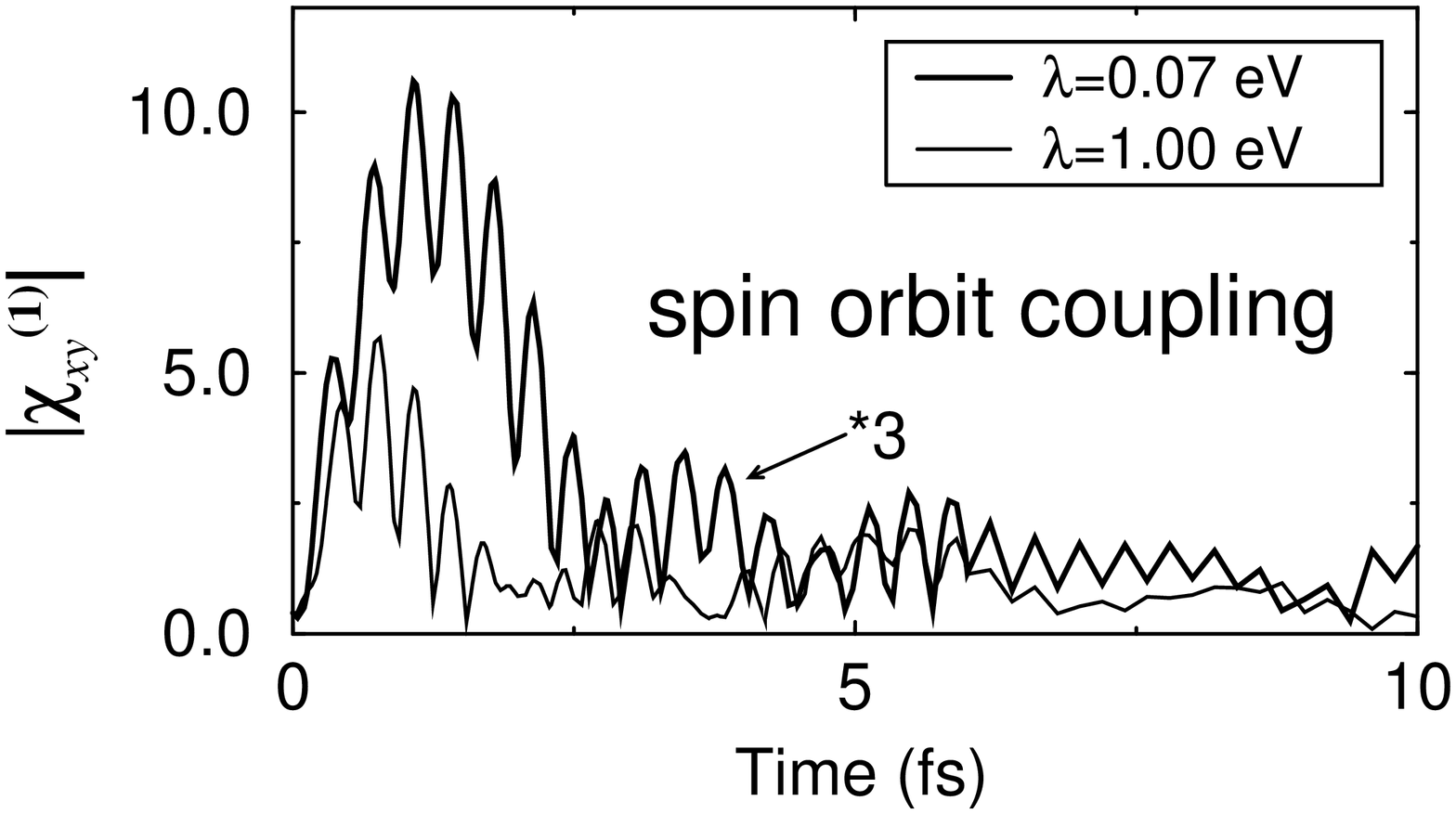,width=8.0cm,angle=270}

{FIG. 2. Effect of spin orbit coupling $\lambda$ on spin 
dynamics. The solid curve is for $\lambda=0.07$ eV while the dashed curve is
for $\lambda=$ 1 eV. SOC may speed up the spin dynamics only in heavy
elements.} 

\psfig{figure=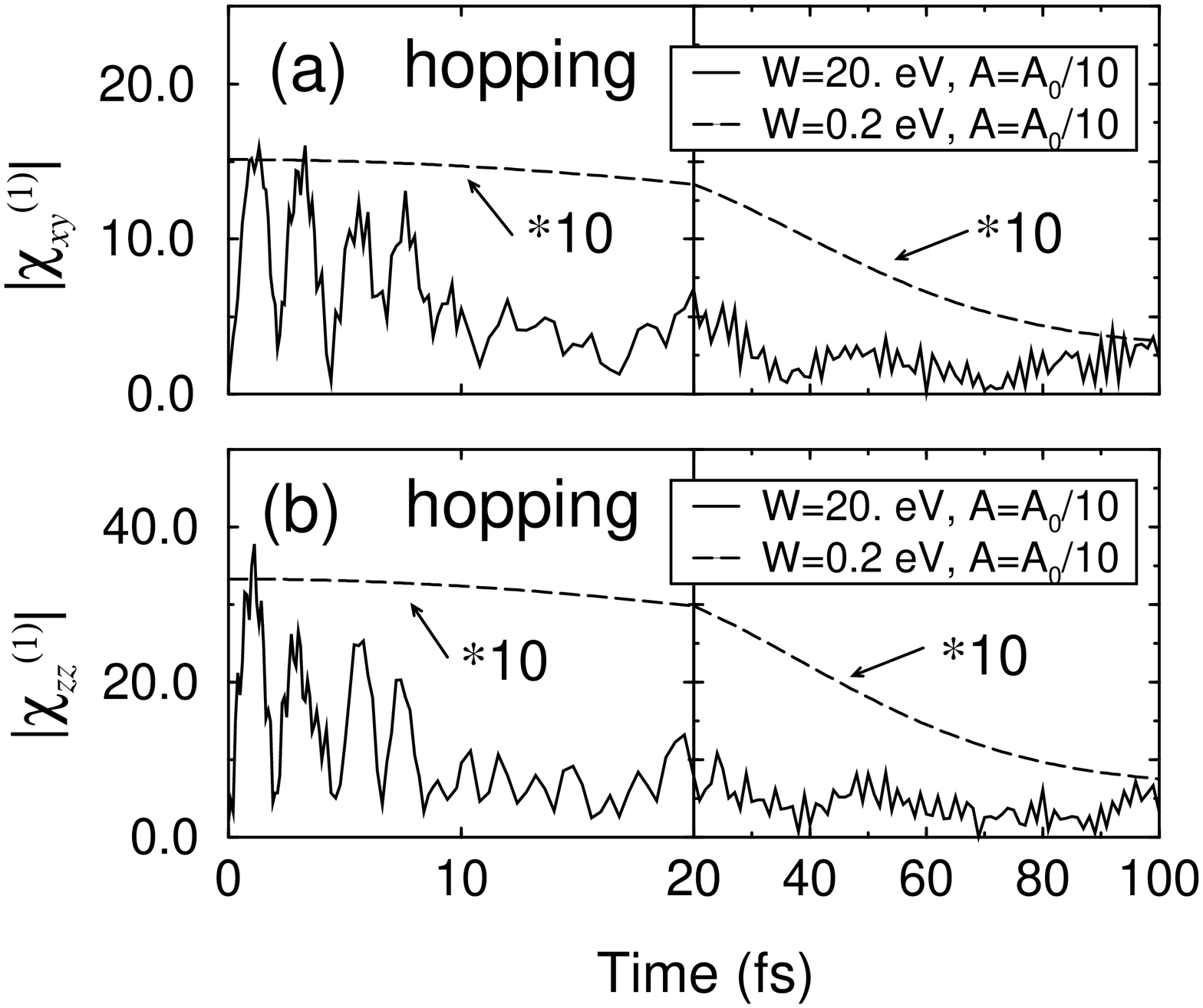,width=8cm,angle=270}

{FIG. 3. Effect of hopping integral on
(a) spin and (b) charge
dynamics. The pulse width
effect is also shown. Nanostructuring and selective population of resonances
slow down the spin and charge dynamics.}

\psfig{figure=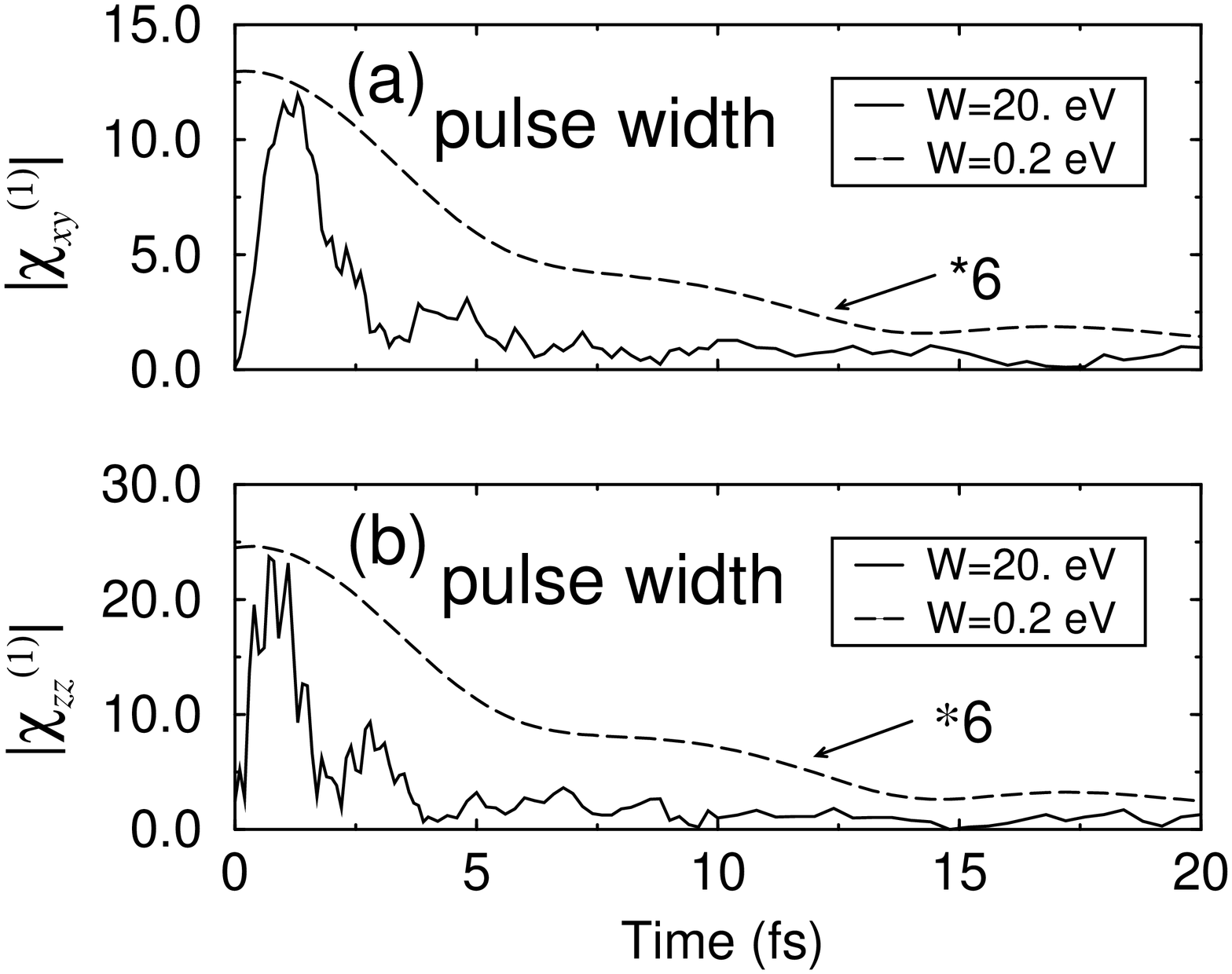,width=8cm,angle=270}

{FIG. 4. Effect of laser pulse-width $W$ on (a) spin and (b) charge dynamics 
for $W=$ 20 and 0.2 eV. Monochromatic laser pulses slow down the dynamics.}
\end{multicols}

\begin{references}
\bibitem{add}A ferromagnetic state is defined to be a state with
multiplicities other than singlets or Kramers doublets.
\bibitem{16}The slow recovery of the spin response during several hundred
picoseconds seen in experiments results from the lattice and is beyond the
scope of the present study.
\end{references}
\end{document}